\documentclass[prodmode,{acmeuroplop}]{acmlarge}
\usepackage{mdframed}
\usepackage{xcolor,colortbl}
\usepackage{array}
\usepackage{caption}
\usepackage{longtable}
\definecolor{gray}{RGB}{106,100,100}

\doi{10.1145/3147704.3147718}

\acmYear{2017}
\acmMonth{7}

\setcopyright{licensedusgovmixed}

\usepackage[ruled]{algorithm2e}
\SetAlFnt{\algofont}
\SetAlCapFnt{\algofont}
\SetAlCapNameFnt{\algofont}
\SetAlCapHSkip{0pt}
\IncMargin{-\parindent}

\newcommand\DOECopyrightNotice[1]{%
  \begingroup
  \renewcommand\thefootnote{}\footnote{#1}%
  \addtocounter{footnote}{-1}%
  \endgroup
}

\title{A Pattern Language for High-Performance Computing Resilience}
\author{Saurabh Hukerikar and Christian Engelmann\affil{Oak Ridge National Laboratory}}

\begin{abstract}
High-performance computing systems (HPC) provide powerful capabilities for modeling, simulation, and data analytics for a broad class of computational problems. They enable extreme performance of the order of quadrillion floating-point arithmetic calculations per second by aggregating the power of millions of compute, memory, networking and storage components.
With the rapidly growing scale and complexity of HPC systems for achieving even greater performance, ensuring their reliable operation in the face of system degradations and failures is a critical challenge.  
System fault events often lead the scientific applications to produce incorrect results, or may even cause their untimely termination.  
The sheer number of components in modern extreme-scale HPC systems and the complex interactions and dependencies among the hardware and software components, the applications, and the physical environment makes the design of practical solutions that support fault resilience a complex undertaking.  
To manage this complexity, we developed a methodology for designing HPC resilience solutions using design patterns. We codified the well-known techniques for handling faults, errors and failures that have been devised, applied and improved upon over the past three decades in the form of design patterns.
In this paper, we present a pattern language to enable a structured approach to the development of HPC resilience solutions. The pattern language reveals the relations among the resilience patterns and provides the means to explore alternative techniques for handling a specific fault model that may have different efficiency and complexity characteristics.
 Using the pattern language enables the design and implementation of comprehensive resilience solutions as a set of interconnected resilience patterns that can be instantiated across layers of the system stack.

\end{abstract}


\acmformat{Saurabh Hukerikar, Christian Engelmann. 2017. A Pattern Language for High-Performance Computing Resilience}

\begin{document}
\begin{bottomstuff}
\end{bottomstuff}

\maketitle

\DOECopyrightNotice{
This work was sponsored by the U.S. Department of Energy's Office of Advanced Scientific Computing Research. This manuscript has been authored by UT-Battelle, LLC under Contract No. DE-AC05-00OR22725 with the U.S. Department of Energy. The United States Government retains and the publisher, by accepting the article for publication, acknowledges that the United States Government retains a non-exclusive, paid-up, irrevocable, world-wide license to publish or reproduce the published form of this manuscript, or allow others to do so, for United States Government purposes. The Department of Energy will provide public access to these results of federally sponsored research in accordance with the DOE Public Access Plan (http://energy.gov/downloads/doe-public-access-plan).}

\section{Introduction}
\label{sec:Introduction}

High-performance computing (HPC) systems, through a combination of massively parallel processing capability and storage capacity, can rapidly solve difficult computational problems in a diverse range of scientific and engineering domains. The design of faster and higher capability HPC systems make significant contributions to scientific progress by providing researchers with capabilities for advanced simulation, computational, mathematical and statistical modeling, visualization and data analytics.
HPC systems are specialized, custom-built machines that are constructed through aggregation of the compute, memory and storage capabilities of hundreds of thousands of components. A typical high-performance computing system contains thousands of processors, several terabytes of memory, and petabytes of storage, and requires highly-customized power and cooling infrastructure.  
The current generation of HPC systems are capable of performing over a quadrillion (10$^{15}$) floating-point arithmetic operations per second.
Yet, there is a significant push by the HPC community to design and deploy next generation of systems that will be capable of exaflops performance (10$^{18}$ operations per second) \cite{Dongarra:2011:IES} in order to enable higher fidelity simulation and predictive analysis capabilities. 

To design, build and effectively operate next-generation exascale-class HPC systems, which will be at least 100 times more capable than the fastest systems today, there are several key challenges that must be addressed \cite{ExascaleTechStudyReport}. Many of these challenges arise from the need to employ hundreds of millions of processing, memory and storage components, and a complex multicomponent software environment to achieve orders of magnitude greater computational performance. 
In addition to the management of unprecedented levels of parallelism, reducing power consumption and coping with an exponentially higher rate of system faults are significant challenges for future HPC system environments.

The challenge of maintaining fault resilient operation is particularly difficult in emerging system architectures that employ hundreds of millions of processing, memory and storage components, and a complex multicomponent software environment, which makes the timely identification and correction of errors much more difficult \cite{Debardeleben:2009}. For long-running simulation, modeling and analysis applications that run on HPC systems, the frequent occurrence of faults may cause incorrect outcomes, or may even lead to fatal crashes of the application program. Therefore, effective resilience solutions that keep HPC applications running to a correct solution in spite of frequent faults will be indispensable for future HPC systems and their applications.  

Solving the resilience problem for extreme-scale HPC systems is a complex undertaking given the growing hardware and software complexity of HPC environments and the emergence of new fault modes at accelerated rates.  
Many of the existing resilience solutions will prove to be insufficient and some will no longer be viable in future systems unless they are significantly reengineered \cite{DARPA:Resilience}. These challenges are elaborated in Section \ref{sec:ResilienceChallenge}. 
To address the resilience challenge, we developed design patterns, which describe the best-known techniques that have been devised and repeatedly applied to confront different types faults, and the resulting errors and failures, in the context of HPC environments over the past three decades \cite{Hukerikar:2017}. The techniques have been formatted as patterns and organized by a hierarchical classification scheme to serve as a resource for designers to draw upon when designing HPC resilience solutions \cite{RDP:Spec}. The design pattern specification describes the detailed descriptions for detection, containment and mitigation of faults, errors and failures that occur in the context of HPC environments. Section \ref{sec:RDP} summarizes the various resilience design patterns and Section \ref{sec:RDP-Classification} presents a classification scheme. 
This paper presents a pattern language for the design and implementation of complete, working HPC resilience solutions. The pattern language, which is introduced in Section \ref{sec:PatternLang}, codifies how the patterns are related to each other. Using the resilience patterns as its elements, the language defines the discipline that makes it possible to combine the individual patterns to create functional resilience solutions. The language enables exploration of pattern-based solutions that have different efficiency and complexity characteristics, and guides a designer from the beginning of a design problem to the successful realization of its solution.
Our pattern language is extremely practical, allowing designers to integrate patterns across the system stack. This pattern language is designed to be useful for HPC hardware and software designers, including system architects, the software developers who implement the libraries and applications for HPC applications, system users and operators.

\section{The High-Performance Computing Resilience Challenge}
\label{sec:ResilienceChallenge}

\subsection{Terminology}
Most modern high-performance computing systems are distributed-memory systems that are architected as \textit{clusters}. They consist of \textit{nodes}, each of which contains processors, memory, and runs its own instance of an operating system. The nodes are connected to each other using high-speed networks. Each node is a shared-memory system consisting of one or more multicore processors; newer node architectures also contain one of more graphical processing units (GPU). Parallel applications distribute their data or tasks across multiple compute nodes to accomplish its work faster. The software environment includes runtimes system frameworks for scheduling, memory management, communication frameworks, performance monitoring tools, numerical libraries, compiler tools, which support and optimize the execution of the parallel application programs. 
The workload of HPC systems consists of scientific and engineering simulation, modeling and analysis programs, that use message passing for exchange of data between processes and synchronization. 
Therefore, in the context of cluster-based HPC systems, and for the purpose of discussing design patterns for such systems, the term \textit{system} refers to an entity that has the notion of a well-defined structure and behavior. A \textit{subsystem} is a set of elements, which is a system itself, and is a component of a larger system, i.e., a system is composed of multiple sub-systems or components. For example, the term system may be used to refer to compute nodes, I/O nodes, network interfaces, disks, etc., or in the software it may be used to refer to a library, runtime framework, or even a function or a variable in a program. The term \textit{full system} refers to the HPC system as a whole, or to a collection of nodes that is capable of running a parallel application. 

In fault tolerance literature, the terms fault, error, or failure have specific meanings. A \textit{fault} is an underlying flaw or defect in a system that has potential to cause problems. A fault can be dormant and can have no effect. When activated during system operation, a fault leads to an error; an \textit{error} event results from the activation of a fault and cause an illegal system state; and a \textit{failure} occurs if an error reaches the service interface of a system, resulting in system behavior that is inconsistent with the system's specification. 

\subsection{Need for Resilience in High Performance Computing Systems}
HPC systems are built using a very large number of nodes each consisting of many processor, memory, network and storage components. With a very large number of aggregate hardware components, the system-level probability that one of them fails is significantly higher. Furthermore, the workload of HPC systems consists of parallel application programs written in languages such as C, C++, Fortran that use a model of message passing between processes. Often the applications use a library implementation of the Message Passing Interface (MPI). However, this flat model of message passing offers no containment for errors or failures; an error in the state of any MPI process may spread to affect the state of processes on other nodes in the system. Similarly, the failure of any one MPI process causes the remaining communicating processes to block indefinitely, which prevents the parallel application from resuming execution.      
The large number of nodes in cluster-based HPC systems with massive number of hardware components and the prevalence of a message-passing-based programming model with inherent failure containment and recovery capabilities, makes resilience an indispensable capability for HPC systems. Resilience solutions are based on a collection of techniques for keeping HPC applications running to a correct solution in a timely and efficient manner despite underlying system faults, errors and failures. 

\subsection{Resilience at Extreme Scale}
In recent years, the progress in the computing capabilities of HPC systems is primarily driven by increasing the number of compute, memory and storage components. These components consist of VLSI chips that are constructed using transistor devices, whose geometries are shrunk every semiconductor process technology generation. As process technology scales further, VLSI devices face new challenges, such as variability, single-event upsets, decreased noise immunity due to NTV operation, transistor device degradation. These effects manifest themselves as unreliable behavior of the components in an HPC system.
In addition to this disturbing trend is the rapidly growing scale and complexity of the hardware and software architecture of modern HPC systems, which makes management of reliability of the system a difficult challenge. The timely detection of faults and degradations, limiting their propagation in the multicomponent system environment, and handling the resulting errors and failures gracefully and efficiently is a daunting challenge for highly-complex, future extreme-scale HPC systems.

\section{Resilience Design Patterns}
\label{sec:RDP}

A resilience solution in a hardware or software component in an HPC system is based on taking appropriate action in the event of a fault, error, or failure.
Many of the techniques for confronting these events that have been devised, applied and improved over the past three decades represent general solutions to recurring problems in the design of resilience solutions for high-performance computing systems. We presented some of these well-known techniques, formatted as resilience design patterns \cite{Hukerikar:2017}. 

Our effort was motivated by the fact that there are a number of hardware and software-based technologies used in HPC systems, but there is a lack of comprehensive methods to facilitate coordination between the hardware and software resilience mechanisms. Often solutions are designed and deployed without fully understanding the protection coverage scope, handling capabilities and efficiency for the different fault models. There are also no established mechanisms and interfaces to connect techniques across layers of the system stack, nor are existing resilience solutions portable to newer HPC system architectures. 

The design patterns are intended to enable structured design and refinement of resilience solutions by using the patterns as building blocks. We believe that the patterns support the design of solutions with a clear understanding of their protection coverage and performance efficiency. Patterns also help in constructing cross-layered resilience solutions that combine capabilities from different layers of the system stack, which effectively balance the performance, resilience, and power consumption. Such systematically designed and well-engineered resilience solutions will be the key to effective and resource-efficient use of the next-generation extreme-scale systems 

The basic, abstract template of a resilience design pattern is defined in an event-driven paradigm, in which each resilience design pattern consists of a {\em behavior} and a set of {\em activation} and {\em response interfaces}. Each pattern describes a problem which occurs on account of a fault, error or failure event, and then describes the core of the solution to that problem. The patterns are written in a prescribed format that describes the problem, solution, design considerations and forces, and the consequences of applying the pattern to various contexts.   
In this remainder of this section, we summarize the various resilience design patterns. The complete pattern catalog, which contains detailed descriptions of the patterns, is available as a specification document \cite{RDP:Spec}.   

\subsection{Strategy Patterns}
The \textbf{strategy} patterns define high-level polices of a resilience solution. Their descriptions are deliberately abstract to enable hardware and software architects to reason about the overall organization of the techniques used and their implications on the full system design. These patterns describe the overall structure of the solution and the key attributes of the solution and their capabilities independent of the layer of system stack and hardware/software architectural features.

\renewcommand{\arraystretch}{2}
\begin{scriptsize}
\begin{longtable}[h]{| m{2.2cm} | m{3.2cm} | m{3.2cm} | m{3.2cm} | m{3.2cm} | }
\captionsetup{width=\linewidth}
\hline
\multicolumn{5}{|c|}{\cellcolor{black}\textcolor{white}{\textbf{Strategy Patterns}}} \\
\hline
\textbf{Pattern Name} & \textbf{Problem} & \textbf{Solution} & \textbf{Forces} & \textbf{Consequences} \\
\hline
FAULT \mbox{TREATMENT} &
The presence of defects or anomalies in a system have the potential to activate, which may potentially lead to an error or a partial/complete failure of the system. &
The pattern attempts to recognize the defect and creates conditions that prevents its activation. The solution requires a monitoring system that observes the key parameters of the monitored system. &
The interactions of the monitoring and monitored systems may interfere with the operation of the system; During the interval for the monitoring system to infer the presence of a fault, it may activate. &
 By preemptively recognizing faults in the system, the pattern prevents their activation, which avoids the need for expensive recovery and/or compensation actions. Requires an additional monitoring system that interferes with system operation. \\
\hline
RECOVERY &
The occurrence of errors or partial/complete failures in an HPC environment prevents applications from running correctly. &
The pattern attempts to recreate the state of the system before the occurrence of an error or failure event. The pattern requires that the system is capable of compartmentalizing and preserving its state for later recovery. &
During error or failure-free operation of the system, the pattern incurs overhead for preserving the system state. &
The pattern handles an error or a failure by substituting an error-free state from the stable storage in place of the erroneous state. Requires periodic creation of recoverable state, which incurs overhead proportional to size of state captured and frequency of state snapshot creation. \\
\hline
COMPENSATION &
Errors or partial/complete failures in an HPC environment cause applications to experience errors or fail. &
The pattern accounts for the error or failure by maintaining sufficient redundancy in the system design. The pattern is based on the definition of modules in a system (with well-defined inputs and outputs), about which redundant information is maintained. &
The pattern introduces a penalty in terms of time (increase in execution time), or space (increase in resources required) independent of whether an error or failure occurs during system operation. &
An error or failure in one of the modules is tolerated by substituting the module with another replica module. The replica must be functionally identical to module it replaces, which incurs cost and/or operation overhead. \\
\hline
\caption{Strategy Patterns}
\end{longtable}
\end{scriptsize}

\subsection{Architectural Patterns}
The \textbf{architectural} patterns convey specific methods necessary for the construction of a resilience solution. They explicitly convey the type of fault, error, or failure event that they handle and provide detail about the key components and connectors that make up the solution. \newpage

\renewcommand{\arraystretch}{2}
\begin{scriptsize}
\begin{longtable}[h]{| m{2.4cm} | m{3.1cm} | m{3.1cm} | m{3.2cm} | m{3.2cm} | }
\captionsetup{width=\linewidth}
\hline
\multicolumn{5}{|c|}{\cellcolor{black}\textcolor{white}{\textbf{Architectural Patterns}}} \\
\hline
\textbf{Pattern Name} & \textbf{Problem} & \textbf{Solution} & \textbf{Forces} & \textbf{Consequences} \\
\hline
FAULT \mbox{DIAGNOSIS} &
An incomplete understanding of the cause and impact of a fault in an HPC system design makes design process of remedial actions difficult. &
The pattern is a derivative of the FAULT TREATMENT pattern and its solution is based on the capability of a monitoring system to analyze the behavior of the monitored system. &
During the time interval for the monitoring system to diagnose the fault, it may activate to cause an error or failure; The degree of accuracy of the fault diagnosis must be high &
The pattern only infers the presence of a defect and reports it, but does act to remedy the fault. \\
\hline
RECONFIGURATION &
The presence of a fault, error or failure event may affect configuration of the system components, preventing its correct operation. &
The pattern is a derivative of the FAULT TREATMENT and the RECOVERY strategy pattern and its solution entails modification of the interconnection between modules in the system as means to prevent activation of a fault, or to recover the system from an error or a failure event. &
The system design must allow for encapsulation of system functions into a set of well-defined modules such that a subset of modules is functionally equivalent to the fault, error, or failure-free version of the system. &
The pattern is based on the encapsulation of system functions into a set of well-defined modules. \\
\hline
CHECKPOINT-RECOVERY &
An unrecoverable error or a failure events in an HPC environment prevents the execution of applications. &
The pattern is derivative of the RECOVERY pattern, whose solution entails maintenance of partial or complete system state on stable storage during error/failure-free operation, or using log-based protocols, which creates a log of non-deterministic events in the system. &
The pattern requires stable storage to capture system state or to log events, which increase overhead in terms of resources required by the system; Incurs error/failure-free overhead to performance &
Upon detection of an error or a failure, the checkpoints/log events are used to recreate last known error/failure-free state of the system before restarting the system. Requires periodic creation of recoverable state, which incurs overhead proportional to size of state captured and frequency of state snapshot creation. \\
\hline
REDUNDANCY &
The occurrence of error or failure events caused by physical faults in an HPC environment prevents the execution of applications. &
Pattern is a derivative of the COMPENSATION pattern; the solution entails creation of multiple redundant versions of a system. The pattern enables a system to tolerate faults that occur because of random phenomena based on the assumption that the random event is unlikely to affect the replicas. &
The pattern introduces penalty in terms of time (increase in execution time), or space (increase in resources required) independent of whether an errors or failure occurs &
The pattern results in a system design consisting of group of N identical replicas of a system's hardware or software components, but there is an implicit assumption of independence of operation between replicas of the system. \\
\hline
DESIGN \mbox{DIVERSITY} &
Design faults introduced by human mistakes or defective design tools cause systems to malfunction or fail &
The pattern is also a derivative of the COMPENSATION pattern, but is based on an approach in which the hardware and software elements for multiple computations are not identical copies, but are independently designed to meet the system's requirements. &
Distinct implementations of the same design specification, which are created by different individuals or teams, incur designer effort and verification costs &
The pattern requires distinct implementations of the same design specification, which are created by different individuals or teams, and with different design tools to systematically avoid design bugs. \\
\hline
\caption{Architectural Patterns}
\end{longtable}
\end{scriptsize}

\subsection{Structural Patterns}
The \textbf{structural} patterns provide concrete descriptions of the solution rather than high-level strategies. They comprise of instructions that may be implemented in hardware/software components. While the strategy and architectural patterns serve to provide designers with a clear overall framework of a solution and the type of events that it can handle, the structural patterns express the details so they can contribute to the development of complete working solutions.

\renewcommand{\arraystretch}{2}
\begin{scriptsize}
\begin{longtable}[h]{| m{2.2cm} | m{3.2cm} | m{3.2cm} | m{3.2cm} | m{3.2cm} | }
\captionsetup{width=\linewidth}
\hline
\multicolumn{5}{|c|}{\cellcolor{black}\textcolor{white}{\textbf{Structural Patterns}}} \\
\hline
\textbf{Pattern Name} & \textbf{Problem} & \textbf{Solution} & \textbf{Forces} & \textbf{Consequences} \\
\hline
MONITORING &
The presence of a defect or anomaly in the system may result in an error or failure &
Derivative pattern of the fault diagnosis architectural pattern. The solution identifies faults based on one of
two strategies: the effect-cause diagnosis, or the cause-effect diagnosis. &
The interactions between the monitoring and monitored systems may interfere with the operation of the system &
The monitoring pattern causes overhead to system operation on account of the additional hardware
or software components required for observation of the system and the cause and effect analysis. \\
\hline
PREDICTION &
Recognizing system conditions that may cause faults may help prevent an error or failure event in the system. &
The solution enables anticipation of fault events using the rule-based method (building rules of association to capture the causal correlations between system parameter values and fault events), or statistical methods (using probabilistic characteristics) to predict the occurrence of future fault events. &
The time interval for prediction must be minimized; degree of accuracy must be high to prevent false positives. &
The prediction adds overhead to system operation, which is related to the complexity of the prediction algorithm. The pattern also incurs overheads on account of actions taken based on incorrect predictions, i.e., false positives and false negatives.  \\
\hline
RESTRUCTURE &
The occurrence of a fault, or a resulting error or failure affects the configuration of a system such that correct system operation is not possible & 
The solution is based on modifying the configuration between the N interconnected subsystems to isolate the subsystem affected by a fault, error or failure. The pattern is a derivative of the RECONFIGURATION architectural pattern. &
The restructuring may cause the system to operate in degraded state using fewer than N sub-systems. &
This pattern seeks to exclude only the affected subsystem from interaction with other subsystems. The resulting system configuration must be functionally equivalent to the system before the occurrence of the event, which is often hard to guarantee. \\
\hline
REJUVENATION &
A fault event, or a resulting error or failure causes a sub-system to operate incorrectly, which prevents correct system operation. &
The pattern isolates the specific part of the system affected by an event and only restores or recreates the affected state with the goal of enabling the system to resume normal operation. The pattern is also a derivative of the RECONFIGURATION pattern. &
The solution requires substantial additional overhead to identify the part of the system affected and perform selective reinitialization. &
The solution requires precise identification of the subsystem affected by a fault, error or failure, and resetting its configuration to guarantee recovery of the overall system.  \\
\hline
REINITIALIZATION &
A fault, error or failure event affects a system to the extent that restoring correct operation is impossible. & 
The pattern performs a reset of the system state to restore pristine state before system operation is resumed. The pattern is also a derivative of the RECONFIGURATION pattern. &
The reinitialization causes loss of all forward progress made by system, but is essential when the effects of an error or failure are unrecoverable. &
The reinitialization is often a slow process, but offers the opportunity to completely remove any effects of the fault, error, or failure. \\
\hline
ROLLBACK &
The occurrence of an error or failure event prevents forward progress of a system. &
The pattern periodically captures system state during regular operation. The pattern may also use log-based protocols for non-deterministic events in the system. The pattern is a derivative of the CHECKPOINT-RECOVERY architectural pattern. Therefore, recovery is performed by restoring the system to the last known stable state. &
Frequent checkpointing increases system execution time, but reduces amount of lost work upon occurrence of an error or failure. &
The pattern introduces overhead during failure-free operation proportional to the size of the system state captured and frequency of checkpointing. However, amount of lost work when failure does occur is inversely related to the frequency of checkpointing. \\
\hline
ROLLFORWARD &
The impact of an error or failure prevents correct operation of a system. & 
Similar to the ROLLBACK pattern, the solution is based on committing system state to persistent storage, or use logging of non-deterministic system events. The defining feature of the pattern is the forward recovery, which restarts operation from the point the system had reached right before the occurrence of the error/failure. &
The post-recovery state of the system created during roll-forward must require minimum recomputation. &
The rollforward operation is often less expensive than rollback algorithm.\\
\hline
FORWARD \mbox{ERROR} \mbox{CORRECTION} \mbox{CODE} &
The presence of information errors in a system's state affects its correct operation. &
The solution offered by the pattern consists of encoding k information symbols that represent the system state and appending a set of r additional symbols. The integrity of the original information and the recovery of any corrupted symbols is performed by decoding the encoded state. The pattern is derived from the REDUNDANCY pattern. &
The strength of the correction code in terms of number of symbols incurs encoding/decoding time and space overhead, but stronger codes provide protection against multi-symbol state errors. &
The encoding and decoding process incurs overhead each time an information symbol is accessed and/or manipulated. The amount of recoverable information depends on the number of additional encoding bits are maintained. \\
\hline
N-MODULAR \mbox{REDUNDANCY} &
The pattern solves the problem of dealing with errors, as well as partial or complete failures. &
The pattern entails creation of a group of N identical replicas of the system. Each of the N replicas may be active simultaneously in various configurations: spatial, temporal, or active on-demand replication. This pattern is also a derivative of the REDUNDANCY architectural pattern &
The n-factor replication of system operation introduces cost and overhead; scope of replication and its inputs, outputs must be carefully selected. &
In order to recover from 2N errors/failures in the system, there must be 2N + 1 distinct replicas of the system. Since the replicas are identical the design effort is low, but overhead of creating N replicas incurs overhead in terms of resources and/or operation time. \\
\hline
N-VERSION \mbox{DESIGN} &
Design bugs may manifest themselves during system operation causing incorrect operation or failure. & 
The pattern entails creation of distinct implementations of the same design specification, which are created by different individuals or teams and with separate design tools. The pattern is a derivative of the DESIGN DIVERSITY pattern. &
The distinct implementation versions of the same design specification must be created by different teams or individuals, and must be verified independently. &
The versions of the system are functionally identical, but designed independently which requires significant amount of design and implementation effort, particularly for complex system specifications. \\
\hline
RECOVERY BLOCK &
Flaws in the design on account of human errors and/or faulty tools may cause errors or failures during system operation &
The solution is based on partitioning the system into distinct functional blocks, in which each block contains at least a primary design and exceptional case handler along with an adjudicator subsystem. This pattern is also a derivative of the DESIGN DIVERSITY pattern. &
The recovery block must be comprehensive in detecting and recovering errors caused by design flaws, without requiring the design complexity or verification effort of a full system design. &
The pattern requires development of a comprehensive acceptance test to validate the result produced by the primary system for various types of design bugs. \\
\hline
\caption{Structural Patterns}
\end{longtable}
\end{scriptsize}

\subsection{State Patterns}
The \textbf{state} patterns describe all aspects of the system structure that are relevant to the forward progress of the system. The correctness and consistency of the system state ensures that the correct operation of the system. These patterns implicitly define the scope of the protection domain that must be covered by a resilience mechanism. The state patterns expose an intrinsic property of the system.

\renewcommand{\arraystretch}{2}
\begin{scriptsize}
\begin{longtable}[h]{| m{2.2cm} | m{3.2cm} | m{3.2cm} | m{3.2cm} | m{3.2cm} | }
\captionsetup{width=\linewidth}
\hline
\multicolumn{5}{|c|}{\cellcolor{black}\textcolor{white}{\textbf{State Patterns}}} \\
\hline
\textbf{Pattern Name} & \textbf{Problem} & \textbf{Solution} & \textbf{Forces} & \textbf{Consequences} \\
\hline
PERSISTENT STATE &
The scope of the system state that remains unchanged for the entire duration of system operation has unique resilience needs from other aspects of the state. &
Encapsulates all aspects of a system’s state that is computed when the system is initialized, but is not modified during the system operation. &
The precise definition of persistent system state requires a detailed understanding of the system structure and operation. &
The encapsulation of such state enables selection of behavior patterns that leverage the persistent property for detection and mitigation of faults and errors. \\
\hline
DYNAMIC STATE &
The scope of the system state that changes as a result of the system operation has unique resilience needs from other aspects of the state. &
Defines the scope of the system state that changes during system operation as the system makes forward progress. &
Precisely scoping the state that changes during operation requires complex analysis. &
The definition of scope of the system state related to system's operation enables the selection of patterns that ensure consistency of the state when performing mitigation actions. \\
\hline
ENVIRONMENT STATE &
The scope of the system state that provide a common set of services that support of primary system function has unique resilience needs from other aspects of the state. & 
Defines the scope of the system state that provides services to the system. &
Separation of the system state that provides a common set of services requires modular design and well-defined abstractions &
The encapsulation of the environment state enables designers to instantiate behavioral patterns that provide detection and mitigation of faults and errors within the supporting services. \\
\hline
STATELESS &
Several resilience strategies operate with the need for a specified protection domain. &
Provides the construct of null state in order to create solutions that have a well-defined notion of behavior but need not define a scope for a protection domain. &
With a stateless pattern, understanding the scope and impact of a resilience solution on a system is difficult. &
By defining the scope of stateless pattern enables instantiation of behavior patterns that do not need to worry about side-effects of their mitigation actions on system state. \\
\hline
\caption{State Patterns}
\end{longtable}
\end{scriptsize}

\section{Classification of Resilience Design Patterns}
\label{sec:RDP-Classification}

\begin{figure}[tp]
  \centering
  \includegraphics[width=\textwidth]
                  {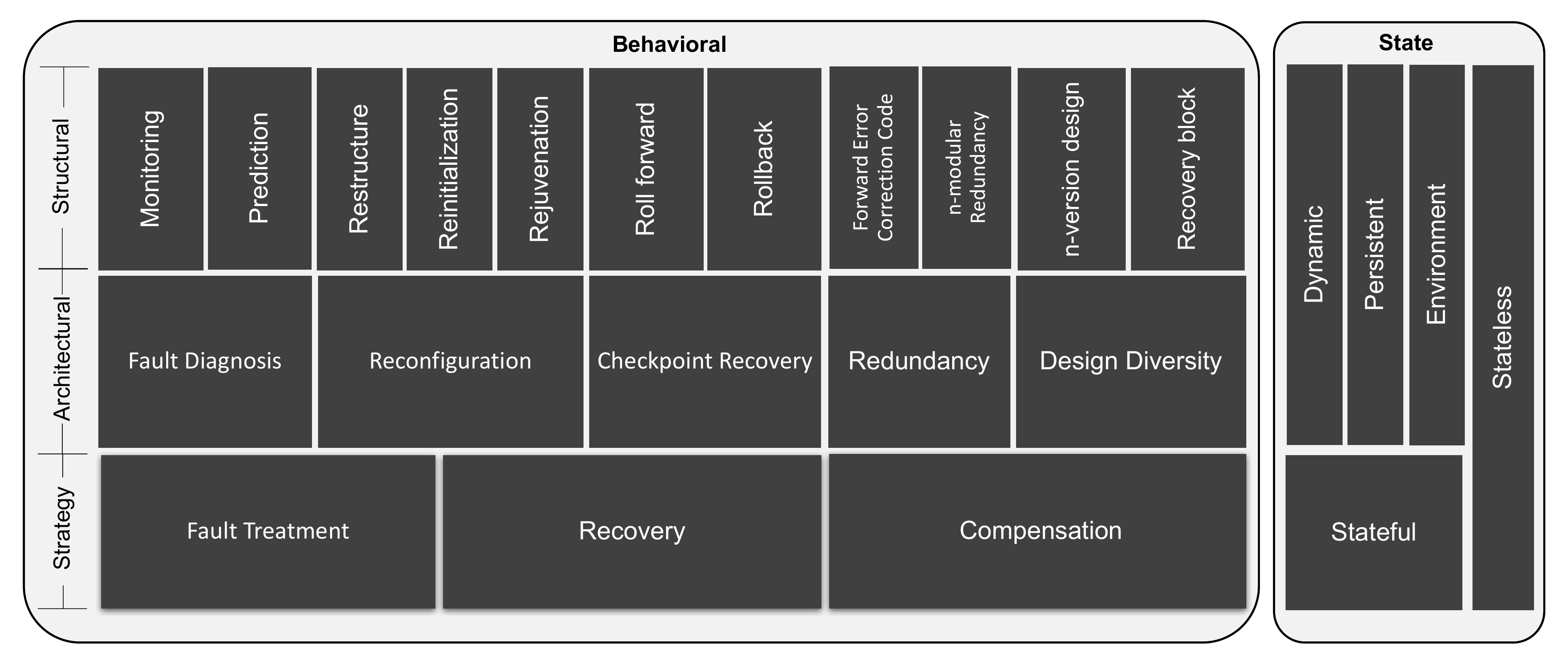}
  \caption{Classification of Resilience Design Patterns}
  \label{Fig:PatternClassification}
\end{figure}

Due to the variety in the granularity and level of system abstraction at which each of the patterns may be implemented, we developed an ad-hoc classification scheme to organize the patterns in our resilience design pattern specification \cite{RDP:Spec}. In this scheme, which is illustrated in Figure \ref{Fig:PatternClassification}, the patterns are organized in a layered hierarchy. 

HPC resilience solutions emphasize the reliability and performance efficiency of an application's execution with the acceptance that the underlying hardware and software environment experiences numerous faults, degradations and component failures \cite{Debardeleben:2009}. Based on this perspective, HPC resilience has two important aspects, namely the forward progress of an application and the consistency and fidelity of an application's data. Accordingly, we have organized the patterns in the catalog into two broad categories: \textbf{state} patterns and \textbf{behavioral} patterns.

The state patterns (described in Section \ref{sec:RDP}) define the protection domain of a resilience solution. These patterns encapsulate particular aspects of a program's state. The different types of state patterns, namely static, dynamic and environment make it possible to define the resilience behavior in a modular fashion; the specific domain scoped by each state pattern may be associated with different resilience techniques. The selection of the state pattern also helps define the containment scope, i.e., the scope of how far a fault or error event propagates. 

The behavioral patterns (described in Section \ref{sec:RDP}) identify detection, containment, mitigation techniques that enable a system, which instantiates and implements these patterns, to cope with the presence of fault, error, or failure events. These patterns are organized in a layered hierarchy that describes the patterns from abstract to concrete descriptions of the techniques. The strategy patterns in the bottom layer are organized by the type of event that they are intended to handle, i.e., whether it is a fault, error or failure. 
The architectural patterns are organized by the specific fault, error or failure types and describe key components and connectors of the pattern solution. The structural patterns in the top layer provide concrete descriptions of the solutions. Their descriptions often contain specific instructions that may be implemented in hardware or software components. 

This classification scheme enables designers to separately reason about scope of the protection domain and the semantics of a resilience pattern's behavior. This hierarchical organization of the behavioral patterns reveals the relationships among these resilience patterns. This classification scheme suggests a number of ways in which these patterns can be combined. Designers may approach the task of developing a complete resilience solution by navigating the hierarchy in top-down or bottom-up manner. This provides the designers with guidelines for selecting patterns for a specific context and combining patterns for the realization of complete resilience solutions. Yet, this scheme leaves much to the skills of the designer since it does not completely cover the various alternative solutions that may exist to address design problems for a particular context. A more comprehensive approach to designing and implementing resilience solutions requires the definition of a pattern language.

\section{Pattern Language for HPC Resilience}
\label{sec:PatternLang}

A pattern language is considered as a system of patterns that are related with each other in a hierarchy or network. The structure of the network helps designers makes sense of the individual patterns, as well as helps anchor them in various combinations to provide complete solutions. 
Our pattern language for HPC resilience explains the discipline to use the various design patterns to create effective and efficient resilience solutions. The elements of the language are the patterns detailed in Section \ref{sec:RDP}. The language guides a designer from the beginning of a design problem to the realization of its solution.

\subsection{Types of Pattern Relations}
In general, a pattern language has the structure of a network such that patterns that are related by some measure of relevance are linked together. The definition of the linkage between patterns is the key for a set of patterns to become a language rather than be seen as a collection of isolated, standalone ideas for design. 

In contrast to a pattern classification, which provides the means to group patterns based on a set of rules or pattern properties, a pattern language explicitly interweaves the patterns in the catalog based on every possible (but at least one) type of pattern interrelation. Based on the interrelations between the patterns, the complete set of the resilience patterns in the catalog forms a language. Therefore, making these relations explicit is essential to the process of developing a pattern language. 

\renewcommand{\arraystretch}{2}
\begin{scriptsize}
\begin{longtable}[h]{ m{3.5cm} | m{7.0cm} | m{3.5cm} } 
\captionsetup{width=\linewidth}
\caption{Types of Pattern Relations} 
\label{tab:pattern-relations}
\endfirsthead
\endhead
\hline
\textbf{Pattern Relation} & \textbf{Description} & \textbf{Inverse Relation} \\ 
\hline
\textit{abstraction}    & Pattern x describes an abstract form of pattern y  & \textit{specialization} \\
\textit{specialization} & Pattern x provides specific details about pattern y & \textit{abstraction} \\
\textit{used with}      & Pattern x is used to address different problem than y; may be used together & \textit{conflict} \\
\textit{conflict}       & Pattern x and y are not suitable to be applied together for a specific problem & \textit{used with} \\
\textit{similarity}     & Pattern x and y have some similar features, but address different problems & \textit{-} \\
\textit{domain}         & Pattern x specifies the protection domain for the behavioral pattern y & \textit{ - } \\
\hline
\end{longtable}
\end{scriptsize}

Highlighting these relationships between patterns enables designers to grasp the entire collection of patterns. Therefore, the pattern language also serves as an index to the catalog of resilience design patterns. For the resilience design patterns, various types of pattern relations may be used to express kinds of relatedness between the patterns. Table \ref{tab:pattern-relations} provides an overview of the types of relationships between the resilience patterns. These interrelations between the patterns form the links between patterns in the network, thereby defining the order in which the patterns should be applied to a HPC resilience design. 

\subsection{Structure of the Pattern Language}
\begin{figure}[tp]
  \centering
  \includegraphics[width=\textwidth]
                  {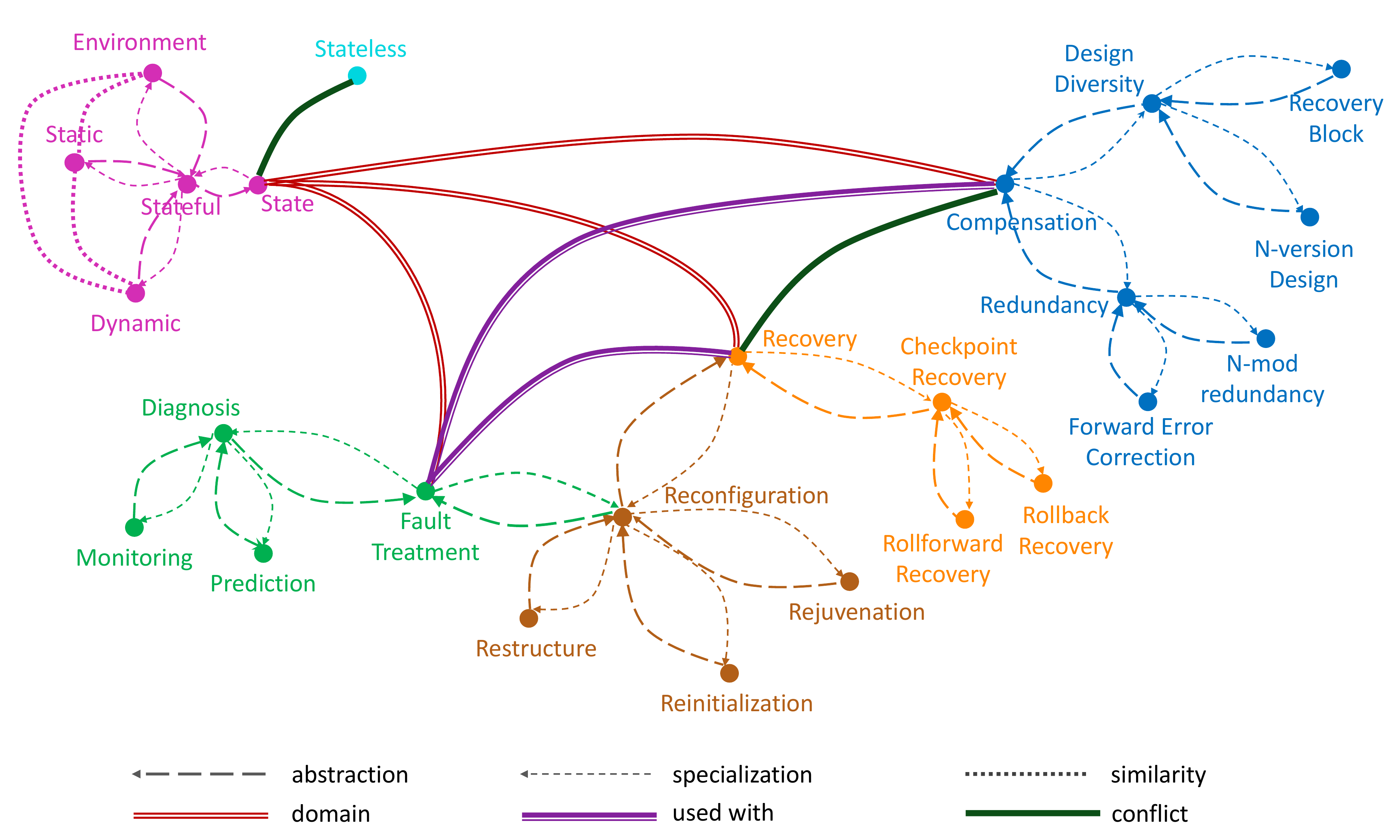}
  \caption{Resilience Pattern Language Representation}
  \label{Fig:PatternLang}
\end{figure}

Forming a pattern language requires establishing rules for linking each of the patterns in the catalog. This is a particularly complex task for resilience design patterns due to their large number and the various design considerations and optimizations that must be accounted for. To enable designers to understand the language and for rapid analysis of the relationships between the various resilience design patterns, we have represented the pattern language using a graph. Each pattern is represented as a vertex and every relation between any two patterns is represented by an edge in the graph network. Based on the type of relation between the patterns, the edges may be directed or undirected. This representation of our pattern language is shown in Figure \ref{Fig:PatternLang}.

The pattern graph represents the language since it captures all the interrelations between the resilience patterns. This representation of the language is intended to make these patterns useful for a broad target audience. System architects may use the language to understand the scope of the problem and develop a high-level layout of the pattern-based solution, while the designers of individual component may use the language to understand the pattern relationships that directly impacts their part of the design.    

The use of the graph representation of the pattern language also enables structured analysis of resilience solutions. For example, a simulator may use the graph representation of the pattern language for design space exploration to evaluate alternative combinations of patterns that may have different complexity and performance characteristics. Similarly, the graph representation of the language may enable a runtime system or scheduler to make dynamic decisions about the suitability of instantiating a specific combination of patterns.   

The graph representation of the language highlights the pattern relations (listed in Table \ref{tab:pattern-relations}) between all the resilience patterns in the catalog. The vertices representing the patterns are clustered to align with the classification scheme described in Section \ref{sec:RDP-Classification}. The state patterns and the three categories of the strategy patterns are represented in different colors. The derivative patterns of each of these classes are represented in the same color as their parents. The patterns are ordered from abstract to concrete to enable designers to focus on the contours of a solution before delving into implementation specifics. Additionally, most of the relations are directed from one pattern to another, but they often also imply an inverse relation in the opposite direction. Therefore, every edge in the graph may be treated as a directed connection between patterns that highlights a specific relation between the two patterns. From the designers' perspective, this representation of the pattern language provides the methodology for selecting patterns from the catalog. The language outlines the ordering of the critical decisions that must be considered when designing and implementing a resilience solution.   

\section{Using the Pattern Language}
Our pattern language spans all the way from the initial architecture of a resilience solution down to the lowest level details of the implementation for a specific architecture and software environment. Defining which patterns to use and how to combine them is the very essence of the pattern language. However, an emphasis of a pattern language is often not represented in the inherent structure of the pattern language. Since our pattern language is in the form of a network, there is no one sequence that perfectly captures the pattern relationships. Therefore, when selecting a suitable combination of patterns for constructing a resilience solution, there are numerous ways in which the network of patterns may be traversed. 

\subsection{Structured Design of HPC Resilience Solutions}
The pattern language outlines the intended flow of information when reading or browsing the pattern catalog. Using the pattern language, solutions are designed incrementally by exploring the links of the network that represents the pattern language. This yields an order in which the patterns should be applied to a design problem, which is called a pattern language sequence. However, the pattern sequence is not strictly linear. Various stakeholders, including system HPC system architects, hardware and software designers, application developers and users can construct solutions by discovering a sequence that fits their design objectives and constraints. For the following key aspects of a resilience design process, the pattern language enables the discovery of pattern-based solutions:  

\begin{itemize}
\item \textbf{Protection Domain:} Based on the scope of the system that the solution intends to protect, the language may traverse the network starting from the state pattern vertices, and then identify the behavioral patterns to protect the selected domain.  
\item \textbf{Fault Model:} The type of event that a solution is designed for forces the designer to consider one of the strategy pattern vertices, before exploring the network links that will enable the identification of derivative patterns that are capable of handling the consequences of a specific fault, error or failure type.
\item \textbf{Fault Management Capabilities:} Based on whether the pattern offers detection, containment, recovery or masking semantics, or a combination of these capabilities, the traversal may commence at specific cluster in the graph representing the language.
\item \textbf{Implementation-Driven:} Often the design of a resilience solution may be constrained by the idiosyncrasies of a hardware architecture or software environment, or by the availability of specific technologies for supporting a resilience solution. In this case, the pattern language may be used to identify the structural patterns first, and traverse the links of the network towards the more abstract behavioral patterns and the state patterns to evaluate the effective protection domain and capabilities.   
\end{itemize}

\subsection{Other Design Considerations for Resilience Solutions}
While the pattern language for designing resilience solutions for HPC systems is intended to provide designers with a roadmap to create solutions, there are various other critical decisions that must be considered in addition to the fundamental choices of protection domain, fault model, capability and implementation mechanisms. These include:  

\begin{itemize}
\item \textbf{Design complexity of the solution:} The effort necessary to incorporate the patterned solution in the overall design of a system.
\item \textbf{Time overhead in the absence of fault, error, or failure events:} The impact of the pattern (in terms of time to solution) on the fault-free operation of a system.
\item \textbf{Time overhead to manage fault, error, or failure events:} The impact on time to solution on account of the actions required to manage an event.
\item \textbf{Space overhead of the solution:} The number of additional components or subsystems that the solution requires.
\item \textbf{Power overhead of the solution:} The impact of applying the pattern on the system’s power consumption.
\end{itemize}

For each optimization objective, the graph edges may be annotated with relations that express the implications of selecting a pattern when traversing the network. Using these additional relations, the pattern language may be used to discover an ordering of patterns that meets these design considerations as well as the functional requirements of a solution for confronting a specific type of fault, error or failure.

\section{Related Work}
\label{sec:RelatedWork}

The solution space for HPC resilience constituted by a number of hardware and software-based solutions is fragmented. HPC vendors have developed a number of hardware resilience technologies, including SECDED ECC for main memory, caches, registers and architectural state, as well as, Chipkill for main memory \cite{Chipkill:Whitepaper:1997}, redundant power supplies and voltage regulators, and reliability, availability and serviceability (RAS) management systems for system-level monitoring and control \cite{Cray:XC40Spec:2014}. On the other hand, various software resilience technologies have been invented, including application- and system-level checkpoint/restart \cite{Mohror:2013:SCR}, fault tolerance extensions to the Message Passing Interface (MPI) \cite{Bland:2013:IJHPCA}, programming models with intrinsic resilience support \cite{Chung:2011:SC}, and algorithm-based fault tolerance \cite{Huang:1984}. However, there hasn't been a concerted effort to develop formal methods for solution space exploration, or for coordination between multiple solutions across the system stack. The resilience design patterns \cite{Hukerikar:2017} and the pattern language presented in this paper are intended to provide HPC designers with an approach to systematically design and implement comprehensive resilience solutions.  

In the context of other types of parallel and distributed systems, the critical need for fault tolerance solutions has driven previous efforts to define design patterns for fault tolerance. For example, patterns have been formalized for the construction of fault tolerance solutions in the context of mission-critical infrastructure, such as telecommunication systems and space programs \cite{Hanmer:2007}. These patterns are intended to offer solutions that meet the stringent reliability requirements of these applications. Fault tolerance patterns that are applicable more generally to various distributed system architectures have also been documented \cite{Saridakis:2002}. These patterns are designed to handle service outages due to crash failures, byzantine failures, omission failures as well as performance failures in distributed systems. Fault tolerance patterns have also been developed in the context of distributed object computing middleware such as CORBA \cite{CORBA:Spec}. These patterns provide support for a range of strategies, including request retry, redirection to an alternative server, passive (primary/backup) replication, and active replication for distributed systems being developed using the standard services and protocols defined by the CORBA standard. While some of the patterns developed for distributed systems are also applicable to HPC systems, certain error and failure modes are unique to HPC environments. The design patterns and pattern language presented in this paper are designed to engineer solutions for HPC environments and focus on optimizing the balance between performance, power and resilience.

\section{Summary}
\label{sec:Summary}

The goal of HPC resilience solutions is to enable effective and resource-efficient use of computing systems at extreme scale in the presence of frequent system degradations and failures. With a new generation of large, heterogeneous HPC systems with multicomponent software environments, the complexity of the system and the interactions between the hardware and software components makes the process of protecting HPC applications from faults and their consequences extremely difficult.  
To navigate this complex landscape, resilience design patterns provide HPC architects and designers with a set of well-known techniques, formatted as patterns, for confronting faults in HPC systems. 
In this paper, we present a pattern language that reveals the relations among the resilience design patterns and provides a discipline for combining the patterns into complete, working solutions. 
The resilience design patterns, and the way they are organized into a pattern language, define a structured approach for architecting practical HPC resilience solutions that address the challenges of extreme rates of fault, error and failures in future HPC systems.

\section*{ACKNOWLEDGMENTS}
This material is based upon work supported by the U.S. Department of Energy, Office of Science, Office of Advanced Scientific Computing Research, program manager Lucy Nowell, under contract number DE-AC05-00OR22725. We thank our shepherd Klaus Marquardt for his comments and suggestions that greatly improved the manuscript. 

\bibliographystyle{ACM-Reference-Format-Journals}
\bibliography{references}


\begin{thebibliography}{00}


\ifx \showCODEN    \undefined \def \showCODEN     #1{\unskip}     \fi
\ifx \showDOI      \undefined \def \showDOI       #1{{\tt DOI:}\penalty0{#1}\ }
  \fi
\ifx \showISBNx    \undefined \def \showISBNx     #1{\unskip}     \fi
\ifx \showISBNxiii \undefined \def \showISBNxiii  #1{\unskip}     \fi
\ifx \showISSN     \undefined \def \showISSN      #1{\unskip}     \fi
\ifx \showLCCN     \undefined \def \showLCCN      #1{\unskip}     \fi
\ifx \shownote     \undefined \def \shownote      #1{#1}          \fi
\ifx \showarticletitle \undefined \def \showarticletitle #1{#1}   \fi
\ifx \showURL      \undefined \def \showURL       #1{#1}          \fi

\bibitem[\protect\citeauthoryear{Bland, Bouteiller, Herault, Bosilca, and
  Dongarra}{Bland et~al\mbox{.}}{2013}]%
        {Bland:2013:IJHPCA}
{Wesley Bland}, {Aurelien Bouteiller}, {Thomas Herault}, {George Bosilca},
  {and} {Jack Dongarra}. 2013.
\newblock \showarticletitle{Post-failure recovery of MPI communication
  capability: Design and rationale}.
\newblock {\em International Journal of High Performance Computing
  Applications\/} {27}, 3 (2013), 244--254.
\newblock


\bibitem[\protect\citeauthoryear{Chung, Lee, Sullivan, Ryoo, Kim, Yoon, Kaplan,
  and Erez}{Chung et~al\mbox{.}}{2012}]%
        {Chung:2011:SC}
{Jinsuk Chung}, {Ikhwan Lee}, {Michael Sullivan}, {Jee~Ho Ryoo}, {Dong~Wan
  Kim}, {Doe~Hyun Yoon}, {Larry Kaplan}, {and} {Mattan Erez}. 2012.
\newblock \showarticletitle{Containment domains: a scalable, efficient, and
  flexible resilience scheme for exascale systems}. In {\em Proceedings of the
  International Conference on High Performance Computing, Networking, Storage
  and Analysis}. 58:1--58:11.
\newblock


\bibitem[\protect\citeauthoryear{DeBardeleben, Laros, Daly, Scott, Engelmann,
  and Harrod}{DeBardeleben et~al\mbox{.}}{2009}]%
        {Debardeleben:2009}
{Nathan DeBardeleben}, {James Laros}, {John~T Daly}, {Stephen~L Scott},
  {Christian Engelmann}, {and} {Bill Harrod}. 2009.
\newblock \showarticletitle{High-end computing resilience: Analysis of issues
  facing the HEC community and path-forward for research and development}.
\newblock {\em Whitepaper\/} (December 2009).
\newblock


\bibitem[\protect\citeauthoryear{Dell}{Dell}{1997}]%
        {Chipkill:Whitepaper:1997}
{Timothy~J. Dell}. 1997.
\newblock {\em A white paper on the benefits of chipkill-correct ECC for PC
  server main memory}.
\newblock {T}echnical {R}eport. IBM Microelectronics Division.
\newblock


\bibitem[\protect\citeauthoryear{Dongarra, Beckman, Moore, Aerts, Aloisio,
  Andre, Barkai, Berthou, Boku, Braunschweig, Cappello, Chapman, Chi,
  Choudhary, Dosanjh, Dunning, Fiore, Geist, Gropp, Harrison, Hereld, Heroux,
  Hoisie, Hotta, Jin, Ishikawa, Johnson, Kale, Kenway, Keyes, Kramer, Labarta,
  Lichnewsky, Lippert, Lucas, Maccabe, Matsuoka, Messina, Michielse, Mohr,
  Mueller, Nagel, Nakashima, Papka, Reed, Sato, Seidel, Shalf, Skinner, Snir,
  Sterling, Stevens, Streitz, Sugar, Sumimoto, Tang, Taylor, Thakur, Trefethen,
  Valero, Van Der~Steen, Vetter, Williams, Wisniewski, and Yelick}{Dongarra
  et~al\mbox{.}}{2011}]%
        {Dongarra:2011:IES}
{Jack Dongarra}, {Pete Beckman}, {Terry Moore}, {Patrick Aerts}, {Giovanni
  Aloisio}, {Jean-Claude Andre}, {David Barkai}, {Jean-Yves Berthou}, {Taisuke
  Boku}, {Bertrand Braunschweig}, {Franck Cappello}, {Barbara Chapman}, {Xuebin
  Chi}, {Alok Choudhary}, {Sudip Dosanjh}, {Thom Dunning}, {Sandro Fiore}, {Al
  Geist}, {Bill Gropp}, {Robert Harrison}, {Mark Hereld}, {Michael Heroux},
  {Adolfy Hoisie}, {Koh Hotta}, {Zhong Jin}, {Yutaka Ishikawa}, {Fred Johnson},
  {Sanjay Kale}, {Richard Kenway}, {David Keyes}, {Bill Kramer}, {Jesus
  Labarta}, {Alain Lichnewsky}, {Thomas Lippert}, {Bob Lucas}, {Barney
  Maccabe}, {Satoshi Matsuoka}, {Paul Messina}, {Peter Michielse}, {Bernd
  Mohr}, {Matthias~S. Mueller}, {Wolfgang~E. Nagel}, {Hiroshi Nakashima},
  {Michael~E Papka}, {Dan Reed}, {Mitsuhisa Sato}, {Ed Seidel}, {John Shalf},
  {David Skinner}, {Marc Snir}, {Thomas Sterling}, {Rick Stevens}, {Fred
  Streitz}, {Bob Sugar}, {Shinji Sumimoto}, {William Tang}, {John Taylor},
  {Rajeev Thakur}, {Anne Trefethen}, {Mateo Valero}, {Aad Van Der~Steen},
  {Jeffrey Vetter}, {Peg Williams}, {Robert Wisniewski}, {and} {Kathy Yelick}.
  2011.
\newblock \showarticletitle{{The International Exascale Software Project
  Roadmap}}.
\newblock {\em International Journal on High Performance Computing
  Applications\/} (February 2011), 3--60.
\newblock


\bibitem[\protect\citeauthoryear{Elnozahy, Bianchini, El-Ghazawi, Fox, Godfrey,
  McKinley, Melhem, Plank, Ranganathan, and Simons}{Elnozahy
  et~al\mbox{.}}{2010}]%
        {DARPA:Resilience}
{E.N.~(Mootaz) Elnozahy}, {Ricardo Bianchini}, {Tarek El-Ghazawi}, {Armando
  Fox}, {Adolfy Godfrey, Forest~Hoisie}, {Kathryn McKinley}, {Rami Melhem},
  {James Plank}, {Partha Ranganathan}, {and} {Josh Simons}. 2010.
\newblock \showarticletitle{System Resilience at Extreme Scale}.
\newblock {\em Whitepaper\/} (2010).
\newblock


\bibitem[\protect\citeauthoryear{Hanmer}{Hanmer}{2007}]%
        {Hanmer:2007}
{Robert Hanmer}. 2007.
\newblock {\em Patterns for Fault Tolerant Software}.
\newblock Wiley Publishing.
\newblock
\showISBNx{0470319798, 9780470319796}


\bibitem[\protect\citeauthoryear{Huang and Abraham}{Huang and Abraham}{1984}]%
        {Huang:1984}
{Kuang-Hua Huang} {and} {J.~A. Abraham}. 1984.
\newblock \showarticletitle{Algorithm-Based Fault Tolerance for Matrix
  Operations}.
\newblock {\it IEEE Trans. Comput.} {C-33}, 6 (June 1984), 518--528.
\newblock


\bibitem[\protect\citeauthoryear{Hukerikar and Engelmann}{Hukerikar and
  Engelmann}{2016}]%
        {RDP:Spec}
{Saurabh Hukerikar} {and} {Christian Engelmann}. 2016.
\newblock {\em Resilience {D}esign {P}atterns: {A} {S}tructured {A}pproach to
  {R}esilience at {E}xtreme {S}cale (Version 1.1)}.
\newblock {T}echnical {R}eport ORNL/TM-2016/767. Oak Ridge National Laboratory,
  Oak Ridge, TN, USA.
\newblock


\bibitem[\protect\citeauthoryear{Hukerikar and Engelmann}{Hukerikar and
  Engelmann}{2017}]%
        {Hukerikar:2017}
{Saurabh Hukerikar} {and} {Christian Engelmann}. 2017.
\newblock \showarticletitle{Resilience {D}esign {P}atterns: {A} {S}tructured
  {A}pproach to {R}esilience at {E}xtreme {S}cale}.
\newblock {\em Supercomputing Frontiers and Innovations\/} {4}, 3 (2017),
  1--38.
\newblock


\bibitem[\protect\citeauthoryear{Inc.}{Inc.}{2014}]%
        {Cray:XC40Spec:2014}
{Cray Inc.} 2014.
\newblock Cray XC40 computing platform.
\newblock   (2014).
\newblock
\showURL{%
\url{http://www.cray.com/Assets/PDF/products/xc/CrayXC40Brochure.pdf}}


\bibitem[\protect\citeauthoryear{Kogge, Bergman, Borkar, Campbell, Carlson,
  Dallya, Denneau, Franzon, Harrod, Hill, Hiller, Karp, Keckler, Klein, Lucas,
  Richards, Scarpelli, Scott, Snavely, Sterling, Williams, and Yelick}{Kogge
  et~al\mbox{.}}{2008}]%
        {ExascaleTechStudyReport}
{Peter Kogge}, {Keren Bergman}, {Shekhar Borkar}, {Dan Campbell}, {William
  Carlson}, {William Dallya}, {Monty Denneau}, {Paul Franzon}, {William
  Harrod}, {Kerry Hill}, {Jon Hiller}, {Sherman Karp}, {Stephen Keckler}, {Dean
  Klein}, {Robert Lucas}, {Mark Richards}, {Al Scarpelli}, {Steven Scott},
  {Allan Snavely}, {Thomas Sterling}, {R.~Stanley Williams}, {and} {Katherine
  Yelick}. 2008.
\newblock {\em ExaScale {C}omputing {S}tudy: {T}echnology {C}hallenges in
  {A}chieving {E}xascale Systems}.
\newblock {T}echnical {R}eport. DARPA.
\newblock


\bibitem[\protect\citeauthoryear{Mohror, Moody, Bronevetsky, and
  de~Supinski}{Mohror et~al\mbox{.}}{2013}]%
        {Mohror:2013:SCR}
{Kathryn Mohror}, {Adam Moody}, {Greg Bronevetsky}, {and} {Bronis~R. de
  Supinski}. 2013.
\newblock \showarticletitle{Detailed Modeling and Evaluation of a Scalable
  Multilevel Checkpointing System}.
\newblock {\em IEEE Transactions on Parallel and Distributed Systems\/}  {99}
  (2013), 1.
\newblock


\bibitem[\protect\citeauthoryear{{Object Management Group}}{{Object Management
  Group}}{2012}]%
        {CORBA:Spec}
{{Object Management Group}}. 2012.
\newblock Common Object Request Broker Architecture (CORBA) Specification,
  Version 3.3.
\newblock   (2012).
\newblock
\showURL{%
\url{http://www.omg.org/spec/CORBA/3.3/}}


\bibitem[\protect\citeauthoryear{Saridakis}{Saridakis}{2002}]%
        {Saridakis:2002}
{Titos Saridakis}. 2002.
\newblock \showarticletitle{A System of Patterns for Fault Tolerance}. In {\em
  Proceedings of 2002 European Conference on Pattern Languages of Programs
  (EuroPLoP)}.
\newblock


\end{thebibliography}

\end{document}